\documentclass[%
 reprint,
superscriptaddress,
frontmatterverbose, 
showpacs,preprintnumbers,
 amsmath,amssymb,
 aps,
]{revtex4-1}
\usepackage[utf8]{inputenc}
\usepackage{graphicx}
\usepackage{dcolumn}
\usepackage{bm}
\usepackage{color,soul}
\usepackage{upgreek}%
\usepackage{hyperref}

\newcommand{\mr}[1]{\mathrm{#1}}

\DeclareRobustCommand{\gobblefour}[4]{}

\newcommand{\um}[0]{~{\upmu\mathrm{m}}}

\newcommand{\ka}[0]{K$\alpha$}
\newcommand{\kaj}[0]{K$\alpha_1$}

\newcommand{\figref}[1]{Fig.~\mbox{\ref{#1}}}

\newcommand{\e}[1]{\times~10^{#1}}

\begin{document}

\preprint{APS/123-QED}


\title{Highly efficient angularly resolving x-ray spectrometer \\
optimized for absorption measurements with collimated sources}

\author{M. Šmíd}
 \email{michal.smid@eli-beams.eu}
\affiliation{Institute of Physics of the ASCR, ELI-Beamlines, 18221 Prague, Czech Republic}

\author{I. Gallardo González}
\affiliation{Department of Physics, Lund University, P.O. Box 118, S-22100 Lund, Sweden}

\author{H. Ekerfelt}
\affiliation{Department of Physics, Lund University, P.O. Box 118, S-22100 Lund, Sweden}

\author{J. Björklund Svensson}
\affiliation{Department of Physics, Lund University, P.O. Box 118, S-22100 Lund, Sweden}

\author{M.~Hansson}
\affiliation{Department of Physics, Lund University, P.O. Box 118, S-22100 Lund, Sweden}

\author{J.C. Wood}
\affiliation{John Adams Institute for Accelerator Science, Imperial College London, SW7 2AZ, United Kingdom}

\author{A. Persson}
\affiliation{Department of Physics, Lund University, P.O. Box 118, S-22100 Lund, Sweden}

\author{S.P.D. Mangles}
\affiliation{John Adams Institute for Accelerator Science, Imperial College London, SW7 2AZ, United Kingdom}
 
\author{O. Lundh}
\affiliation{Department of Physics, Lund University, P.O. Box 118, S-22100 Lund, Sweden}
 
\author{K. Falk}
\affiliation{Institute of Physics of the ASCR, ELI-Beamlines, 18221 Prague, Czech Republic}

\date{\today}

\begin{abstract}

Highly collimated betatron radiation from a laser wakefield accelerator is a promising tool for spectroscopic measurements. Therefore there is a requirement to create spectrometers suited to the unique properties of such a source.
We demonstrate a spectrometer which achieves an energy resolution of \textless 5 eV at 9 keV ($E/\Delta E>1800$) and is angularly resolving the x-ray emission allowing the reference and spectrum to be recorded at the same time. The single photon analysis is used to significantly reduce the background noise. Theoretical performance of various configurations of the spectrometer is calculated by a ray-tracing algorithm. 
The properties and performance of the spectrometer including the angular and spectral resolution are demonstrated experimentally on absorption above the K-edge of a Cu foil backlit by laser-produced betatron radiation x-ray beam.
\end{abstract}

\maketitle

\section{Introduction}
Betatron radiation generated by a Laser Wake Field Accelerator (LWFA) \cite{rousse,kneip2010} is an x-ray source producing broadband radiation with synchrotron-like spectrum coming in ultrashort ($< 30$ fs duration) and highly collimated beams ($\approx$ 10-20 mrad). Its unique qualities make it an ideal probe for active time resolved diagnostics of ultrafast processes in mid-Z elements \cite{Albert}. It could be used for x-ray absorption spectroscopy, most likely for observation of absorption lines to infer the ionization degree and temperature, or for the x-ray absorption near-edge spectroscopy (XANES) measurement to investigate the electronic density of states and therefore ionic correlation and structure of the matter \cite{rehr}. However, the generally low flux of the beam produced using lasers with moderate power makes it very challenging for detection. Thus a highly efficient spectrometer is necessary for such applications.

XANES is a diagnostic method investigating the x-ray absorption just above the K-edge, i.e. capturing of x-rays in the process of photo-ionization of the $1s$ electrons. These electrons are ejected into the free (continuum) state with low residual energy and they undergo scattering by surrounding ions. Therefore, the absorption cross-section reflects the ionic structure of the matter \cite{rehr}. Time resolved studies of K-edge in laser shocked Al with a 10 ps long backlighter have been used to determine the Warm Dense Matter (WDM) temperature \cite{dorchies}. The electronic structure during the creation of warm dense molybdenum was studied via the XANES spectroscopy and compared to theoretical calculations \cite{dorchiesMo}.
XANES of Fe can be used to investigate hydrodynamic conditions during shock compression \cite{harmand}. 
The first use of laser-driven betatron radiation beam as a backlighter for absorption studies did not provide sufficient number of detected photons \cite{mo}.
This was caused mainly by a significant loss of photons during their refocusing and lower efficiency of the spectrometer compared to this work.

In this article we present a design for a crystal spectrometer optimized for highly efficient detection of this kind of radiation. This spectrometer employs a mosaic Highly Oriented Pyrolitic Graphite (HOPG) crystal which guarantees extremely high reflectivity compared to more often used monocrystals. Though these crystals usually produce lower resolution spectrum, we have achieved a sufficient resolution for most diagnostic purposes mainly due to longer crystal to detector distance and by minimizing geometrical effects. A defocused regime of the von Hamos setup is proposed that provides angular resolution to resolve a probe and a reference beam spectrum at the same time.

\begin{figure}[bh]
\includegraphics[width=8cm]{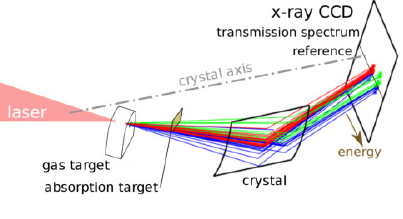}
\caption{\label{spectrometer} Schematic of the spectrometer setup. The red and green lines denote x-rays going through the absorption target with two different energies. Both energies of x-rays propagating next to the target are marked in blue color and constitute the reference beam.}
\end{figure}

A spectrometer design for LWFA betatron radiation has two significant requirements: high efficiency and suitability for highly collimated beams. Mosaic crystals are often used for high collection efficiency setups, however they are usually placed close to the source to cover large solid angles. This close proximity to source negatively affects the spectral resolution and signal to background noise ratio. 
The collimation of the betatron beam allows the relatively small crystal to be placed further away from the source while maintaining overall collection efficiency and low geometric aberrations, common in larger focusing crystals.

The mosaic crystals are composed of a set of small crystallites whose surface is slightly inclined from the crystal surface. The crystals are characterized by their mosaic spread, which is the FWHM (full width at half maximum) of the angular distribution of the crystallites. The rocking curve and topography of large ($5\times10$ cm) thin mosaic crystal in a similar configuration has previously been studied in detail, and these crystals were used for x-ray Thomson scattering measurements \cite{glenzer,zastrau,Zastrau2013}. This kind of measurement is characterized by the need of high energy resolution and by the use of low intensity divergent sources. The spectral resolution was limited by focal aberrations due to the large collection angles. The HOPG crystal based spectrometers have also been used to obtain the angular resolution with divergent sources \cite{Saiz}.

The spectral resolution reported in previous works with this type of crystal is usually not so great \cite{legall2009,zastrau}. Resolution of 8 eV FWHM was reported on similar crystal at 7.5 keV \cite{Ice}. 
By increasing the crystal to detector distance \cite{legall2009} and exploiting the low Bragg angle in this configuration it was possible to significantly improve the spectral resolution compared to previous work.

In the current article the performance of the spectrometer is predicted for various configurations by the ray-tracing simulations. The most suitable configuration was experimentally demonstrated and evaluated by measuring absorption spectra using a betatron radiation source. The range of the measured spectra extends up till $\approx 150$~eV above the edge, therefore it contains both the XANES and EXAFS regions \cite{rehr}.
The experimental measurement of the spectral Point Spread Function (PSF) is shown. The spectral resolution and penetration depth of the radiation into the crystal is inferred from this measurement.

\section{Spectrometer design}
This section describes the principles and parameters of the spectrometer. It shows its basic properties like angular resolution and response function and it presents the ray-tracing simulations and discusses the theoretical performance of the spectrometer.
\subsection{Crystal}
The crystal used was commercially available from the Optigraph company \cite{optigraph}. The mosaicity grade was ZYB with mosaic spread $m=0.8^\circ$. The size was $20\times20$ mm and its surface was cylindrically bent. The radius of the curvature was obtained by measuring the position of the focus of a divergent laser radiation reflected by the crystal as $r=108\pm5$~mm. The uncertainty is so high because the focus is several millimeters large due to the irregularities of the crystal surface which are common in HOPG crystals. The nominal radius given by the manufacturer is 115 mm. 
 
\subsection{Geometry}
The presented spectrometer is based on the well known von Hamos geometry \cite{vonhamos}, as illustrated in \figref{spectrometer}. The cylindrically curved crystal defines the axis of the cylinder. If the focused setup is used, both source and detector are located on this axis in opposite directions from the crystal. The radiation coming from the source is focused to the detector in one direction, forming a line with dispersion along its length. The dispersion is governed by various Bragg angles of reflection of different x-ray energies.
However, due to the use of mosaic crystal, several significant differences to the scheme arise.

In a von Hamos scheme with a monocrystal, there is no focusing of the radiation along the dispersion direction, while in the perpendicular direction, both source and detector have to be placed on the center of the crystal curvature to obtain spatial focusing.
When the mosaic crystal is used, the effect of \emph{mosaic focusing} \cite{legall2009} focuses the radiation in the spectral direction like in the Rowland circle scheme. This means that the crystal works in a similar way as if it was toroidally bent with a large radius of curvature in the dispersion plane. 
On the other hand, the mosaicity includes randomness in the reflection angle. Therefore the focusing in the spatial direction is not perfect.

We can define the ratio $r_\mr{m}$ for mosaic crystal spectrometers which shows the relative importance of the mosaicity effects:
\begin{equation}
 r_\mr{m} = \frac{m}{ \Delta \theta} \approx \frac{m l}{l_\mr{c} \sin(\theta_0)}
\end{equation}
In this formula, $m$ is the mosaic spread of the crystal, $\Delta \theta$ is the maximal difference between angle of incidence of incoming radiation on various places on the crystal, $l$ is the source to crystal distance, $\theta_0$ is the central angle of incidence, and $l_\mr{c}$ is the crystal length.

If this ratio is high enough ($r_\mr{m} \gg 1$) the mosaicity effects dominate over the variation of angle of incidence over the crystal surface.  This has two important consequences. First, the whole surface of the crystal reflects basically the same energies of radiation. This means that a ray with given energy can be reflected at any position within the crystal surface. However, the mosaic focusing guarantees that the photon is reflected to an angle based on its energy, therefore impinging the detector at position given by the dispersion relation. Second, the spectral range is governed by the mosaicity, not by the variation of the incidence angle on the crystal (size of the crystal). 

 It is important to note that $\Delta \theta$ is minimum of beam divergence and crystal solid angle. 
 Therefore the ratio can never be small enough and the spectral range is usually governed by the mosaicity for highly collimated sources like LWFA betatron beam.
In the current setup, $m=0.8^\circ$, $l=700$~mm, $l_\mr{c} = 20$~mm and $\theta_0=11.88^\circ$ therefore $\Delta \theta \approx 0.34^\circ$  and the ratio was $r_\mr{m} = 2.4$.






\subsection{Angular resolution}
\begin{figure}[ht]
\includegraphics[width=4cm]{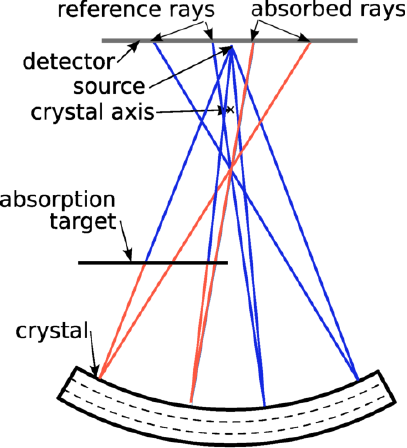}
\caption{\label{angular} Demonstration of the angular resolution. All rays (initially in blue color) are coming from a point source, half of them propagates through the absorption target (they become drawn with red color). Those rays hit the right side of the detector.}
\end{figure}

If both the source and the detector would be located on the axis of the cylindrical crystal, any radiation coming from a point source could be reflected to a thin spectrum on the detector. Any radiation emerging from a source positioned slightly off the axis would be reflected to a shifted position, giving rise to the typical \emph{spatial} resolution. 
Since for the betatron backlit measurements we are interested in resolving different rays going from the same point-like source this type of resolution is not useful. 
On the contrary, \emph{angular} resolution is needed because 
it is resolving the x-rays passing through the sample from the reference part of the beam, see \figref{angular}.
The angular resolution with this type of spectrometer can be achieved when both the source and detector are located off the crystal axis, as illustrated in Figs. \ref{spectrometer} and \ref{angular}. The image of a point source is defocused and each position on the detector corresponds to a different emission angle from the source.

Mosaicity has a negative effect on this resolution.
In the focused setup with a perfect monocrystal the width of the signal would correspond to the source size. The random orientation of the crystallites blurs this signal and increases its size to several millimeters, see the horizontal spread (green) in \figref{setups}.
This effect significantly degrades the angular resolution, but it is beneficial since it improves the dynamic range for single photon counting techniques as is discussed in greater detail in 
Sec.\ref{singlephoton}. 

\subsection{Ray-tracing}
A new Monte Carlo ray-tracing code with the mosaicity effects included has been developed to model the properties of the spectrometer. The code is similar to the one developed by Zastrau \emph{et al.} \cite{zastrau}, though the handling of radiation penetration through the crystal is different, as described below.
This code generates a set of x-rays with random initial directions, lets them interact with the crystal and tracks their intersection with the defined detector plane. The interaction with the crystal is defined by two important functions defining the properties of the crystal: \emph{penetration depth distribution} of the photons and the \emph{mosaic spread function}, i.e. the angular distribution of the crystallites.
The code selects the depth where each ray is reflected based on the penetration depth distribution. This behavior simulates that the photon has to travel some unknown distance through the crystal until it finds the crystallite with appropriate orientation to be reflected. This distribution follows the exponential decay $e^{-\mu x}$ to mimic the absorption of x-rays in the matter. The absorption length $1/\mu$ is generally not known. It is smaller compared to absorption in common graphite forms due to the structure of the crystal \cite{legall2009} and it has to be inferred from experiments.

In the second step, the code randomly selects the orientation of the crystallite following the \emph{mosaic spread function } and in a way that it fulfills the Bragg condition, which defines the angle between the crystallite and the ray. Finally, the angle of the reflected ray is calculated and the intersection with detector is found.
From a typical set of usually $10^{5}$ photons, the synthetic detected image is formed and analyzed to provide the spectrometer characteristics described in the following section.

\subsection{Theoretical performance}

The main aim of the ray-tracing simulations is to predict the performance of the spectrometer and thus to help with finding an optimal setup for given experimental requirements. The feasibility of the spectrometer for various x-ray energies was studied. Since the setup with Cu at 9 keV was found as suitable, the performance of crystal with various radii was investigated on this energy. As the energy and crystal radius was chosen, the effect of the distance $l$ is studied.
There are five performance parameters tracked by the ray-tracing simulations:
\begin{itemize}
 \item \emph{Spectral resolution [eV]}: the FWHM of the spectral point spread function (the simulation uses monochromatic signal).
 \item \emph{Energy range [eV]}: the interval between the most and less energetic rays reflected by the crystal. Note that the efficiency is significantly dropping close to the edges of the range, thus the usable range is approximately half of this value.
 \item \emph{Angular resolution [mrad]}: horizontal FWHM of a perfectly collimated beam.
 \item \emph{Horizontal spread [mm]}: the FWHM of the signal trace of a divergent beam in the direction of the angular resolution (perpendicular to the spectral one).
 \item \emph{Efficiency [\%]}: Ratio of rays reflected by the crystal compared to all rays in a monoenergetic beam collimated to 20 mrad.
\end{itemize}

\begin{figure}[h!]
\includegraphics[width=7.05cm]{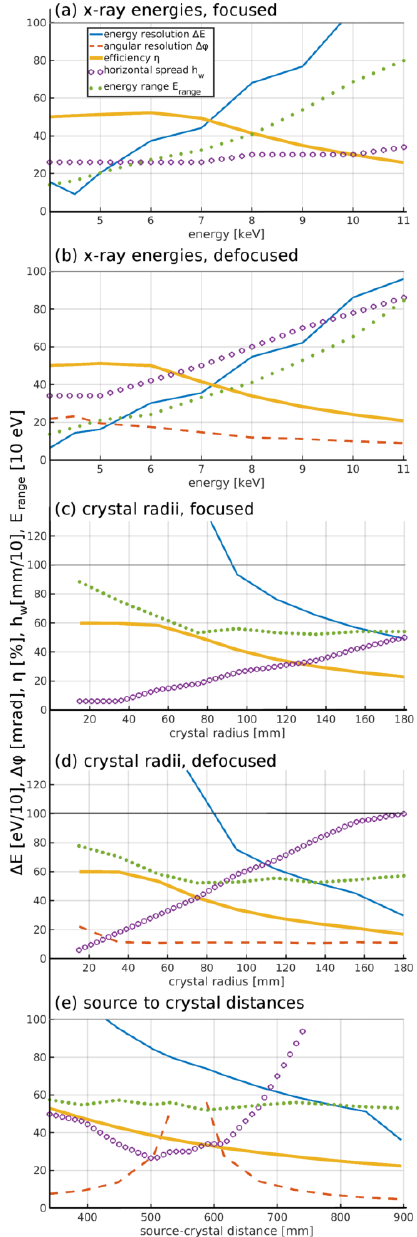}
\caption{\label{setups} Ray-tracing predicted performance of the spectrometer for various x-ray energies (a,b), crystal radii (c,d), and source--crystal distances (e). Calculated either for the focused setup (a,c) or defocused by 25\%.}
\end{figure}

Figure \ref{setups} a) shows the variation of these parameters when the spectrometer is setup for different x-ray energies. The crystal dimensions are $2\times2$ cm, thickness 2 mm and radius of curvature is set to $r=115$ mm in the simulations. In general, HOPG crystals are suitable for range $2-11$ keV, however the setups with energies below 4 keV are considered unfeasible due to the increasing Bragg angle. 
The energy range is significantly dropping with decreasing energy, being only 180 eV at 4 keV. 
From Bragg law it can be derived that the variation of energy for fixed variation of Bragg angle is close to $E_\mr{range} \propto E^2$.
The efficiency of the spectrometer is slightly decreasing with energy because the crystal is located further from the source making the effective solid angle smaller. 
The energy resolution is getting worse with increasing energy. 
The energy smearing by the effect of penetration depth is the dominant one. This effect gets stronger with increasing x-ray energy due to higher penetration depth, therefore the resolution gets worse with energy (\figref{setups} a,b). On the other hand, the influence of this effect decreases with increasing crystal to detector distance, therefore the spectral resolution improves with distance (\figref{setups} c,d,e).


The simulations in \figref{setups} a) and c) were calculated in the focused geometry therefore the angular resolution is not present.  \figref{setups} b) and d) show the configurations in the defocused regime. This was achieved by increasing both the source to crystal and crystal to detector distance by 25\%. The angular resolution of approximately $10-30$ mrad is obtained, the horizontal spread of the signal is more than doubled compared to its focused value and the spectral resolution get slightly better due to the increased distances.

Figures \ref{setups} c) and d) show setups for $E=9$ keV and variable crystal radii for the focused and defocused regimes respectively. The source to crystal distance is increasing linearly with the crystal radius ($L=r/\sin(\theta)$) which improves the energy resolution ($\Delta E \propto 1/L \propto 1/r$), increases the horizontal spread ($h_w=2\sin(m)r$), and decreases the efficiency.

The performance of the setup for 9 keV radiation, $r=115 ~\mathrm{mm}$, and various source to crystal distances is shown in \figref{setups} e). The geometry becomes focused when both source and detector are located on the axis of the crystals cylinder, which happens at $l=560$ mm. The regime defocused by 25\% in distance
has $l=700 $ mm.
The angular resolution is improving with increasing distance from the focus. There is no angular resolution close the focused position. The focus is clearly seen as the minimum of the horizontal spread. The value of this minimum is given by the mosaicity of the crystal. The fact that the horizontal spread significantly increases for large distances to more than 10 mm is usually considered to be undesirable.  However, this can reduce the flux on the detector to such values that the single photon counting method can be used, see Sec. \ref{singlephoton}.
Similarly as in \figref{setups} a) -- d), the efficiency is decreasing with distance due to smaller observation angle while the resolution improves.
This better spectral and angular resolutions make the long distance setups more favorable, which is also beneficial in the experiment since longer distance significantly lowers background on the detector caused by the interaction of the accelerated electrons with surrounding material. The upper constraint for the distance is then given by the size of the detector which was 27 mm in our case. However, the limited size of the experimental chambers make these setups very challenging and may limit their use.


\subsection{Response function measurement on Cu \ka}
The response function of the spectrometer was tested on a Cu \ka ~source. 
The x-ray source was a microfocus tube producing the Cu \ka ~doublet at $E=8027$ and 8047~eV. It was running at power 4 W and produced a source size of $20 \um$.
The \kaj ~is narrow enough to be used as a quasimonochromatic source to evaluate the spectral resolution.

\begin{figure}[ht]
\includegraphics[width=8cm]{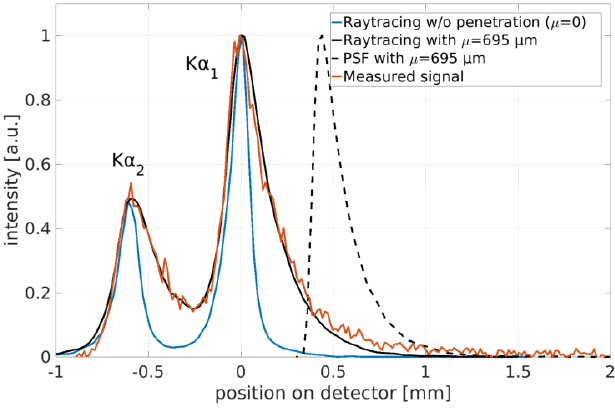}
\caption{\label{rigaku} Measured spectrum of Cu \ka~compared with results of ray-tracing. The dashed line shows the calculated PSF of the spectrometer (drawn with horizontal offset)}
\end{figure}

Figure \ref{rigaku} shows the experimentally measured \ka ~profile ~ (red) alongside results of the ray-tracing simulations. The blue line denotes a result when the penetration depth is neglected, i.e. all rays are expected to reflect from the crystal surface. The solid black line corresponds to a situation when the penetration depth $1/\mu$ was set to $695 \um$ to make the best fit with the experimental data. The dashed line presents the point spread function (PSF) of the spectrometer, i.e. the image of a monochromatic source, when this value is used in the ray-tracing. 
The assymetric shape of this PSF reflects the penetration depth broadening \cite{pak}. The peak of the curve consists of the rays reflected from the surface of the crystal. The deeper the rays are reflected, the more they are shifted to the higher energy part of the PSF due to their extended trajectory in the crystal. This explains the sharpness of the left boundary and the exponential decay towards the right.
The FWHM of this curve is 5.9 eV.

This penetration depth $1/\mu = 695 \um$ is a fixed parameter of the crystal and photon energy and can be used for the ray-tracing simulations. Knowledge of the PSF is crucial for evaluation of the data. The predicted spectra should be convolved with this PSF prior to comparison with the measured data. Since this PSF is strongly asymmetrical the convolution with a Gaussian or Lorentzian function usually used for this purpose would not lead to a satisfactory agreement.

\subsection{Efficiency}
The efficiency calculated by the ray-tracing code is defined as a ratio of photons hitting the detector to the total number of photons at the central energy within the 20 mrad divergent cone. This efficiency is determined to be 30\% for the setup used in the experiment described further below (photon energy 9 keV, crystal radius of curvature 115 mm and source to crystal length 700 mm). The code expects that all photons which hit the crystal with suitable angle are reflected from it, therefore this number has to be multiplied by the crystal peak reflectivity which is 52\% \cite{freund}. The quantum efficiency of the CCD used in current experiment was 14\% and the total transmission of used filters was 75\%. 
The total efficiency is obtained by multiplying all these coefficients as 1.6\%. This  number relates the number of detected photons with certain energy to all photons with given energy produced by the x-ray source, assuming it has 20 mrad divergence.

\section{Experimental results}
\begin{figure}[t]
\includegraphics[width=8cm]{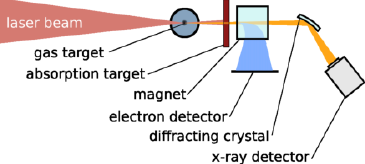}
\caption{\label{scheme} A schematic of the experiment. The laser is focused to the gas target where the electron and x-ray beams are generated. The electron beam is deflected by the magnet to the scintillator. The x-ray beam propagates through the investigated absorption target is diffracted by the crystal and detected by the CCD.}
\end{figure}

The spectrometer was tested at the Lund Laser Centre using a multi terawatt laser operating at a central
wavelength of 800 nm. The laser system produces 37~fs long pulses of up to
800 mJ after compression. The shape of the focal spot was optimized using deformable mirror to a size of $\approx 14 \um$ FWHM.
The experimental setup is illustrated in figure \ref{scheme}. The laser pulse was focused onto an entrance of a 6 mm long 99\% He and 1\% N$_2$ mixture gas cell with a 
775 mm focal length off-axis parabola. The electron density in the gas cell is $1\e{19}~\mathrm{cm}^{-3} $. During the interaction the laser drove a nonlinear plasma wave. The electric fields inside this wave accelerated electrons to $\approx 200~ \mathrm{MeV}$ and the oscillatory movement of these electrons produced the betatron radiation which was collimated to $\approx 20$ mrad and had a source size $\approx 10\um$. The critical energy of the x-ray spectrum was estimated as 2 keV by analysis of the transmission through Ross filters \cite{desforges}.

A permanent magnet was placed after the gas cell to deflect the electrons to protect the crystal as well as to be used as electron spectrometer. The HOPG crystal was placed further down the x-ray beam to reflect the x-rays onto a direct detection Princeton Instruments PI-MTE CCD with chip size $27\times27$ mm and pixel size $13.5\um$. The absorption target was placed 3 cm behind the gas cell (before the magnet) in order to minimize the probed region. The spectrometer was set up for the Bragg angle $\theta = 11.88\pm0.07^\circ$ corresponding to the Cu K-edge at 8980~eV. The source to crystal distance was set to $l=697\pm3~\mathrm{mm}$ which is about 125\% of the the distance for focused geometry. The crystal was located on a motorized rotational stage in order to achieve high precision alignment. 

Following sections draw three important conclusions about the spectrometer: the possibility of noise removal via single photon counting, the usability of angular resolution, and the evaluation of spectral resolution.

\subsection{Single photon counting analysis}
\label{singlephoton}

The large horizontal spread of the signal on the detector and  the relatively low photon flux allowed to use the single photon counting regime \cite{Higginbotham}.
The signal of typically 33000 photons per shot on the CCD was stretched over $\approx 2 ~\rm{cm}^2$ which corresponds to $\approx $ million pixels. Therefore there were about 1 photon per 30 pixels which is low enough resolve single events. Higher fluxes might cause too many event overlapping making the data analysis more complicated.

\begin{figure}[t]
\includegraphics[width=\columnwidth]{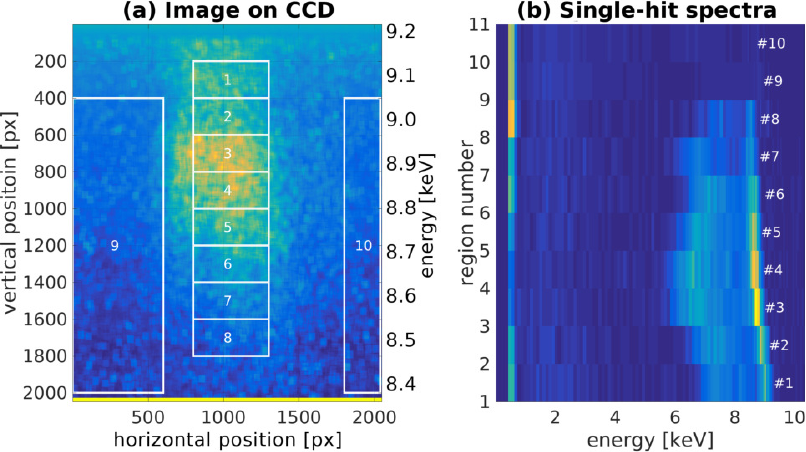}
\caption{\label{single} Raw single shot data from the CCD with several selected regions (a) and x-ray spectra for those regions based on single photon counting (b). The shift of peak around 9~keV shows the dispersion of the spectrometer. Signal below 5~keV comes from experimental background.}
\end{figure}

There are ten regions marked in a single shot raw data in figure \ref{single} a). Regions 1 -- 8 correspond to part of the spectrum with various x-ray energies while regions 9 and 10 are areas without signal, having only background noise and scattered radiation. Figure \ref{single} b) shows the broadband spectrum for each region obtained by the single photon counting analysis.
All regions with data show a dominant peak around 9 keV. The dispersion of the spectrometer is clearly seen in the decrease of this peak energy with increasing index of the region. Regions 9 and 10 show almost no x-rays at this energy, while the low energy signal (below the threshold of 0.5 keV) and a broad peak around 2 keV is present almost homogeneously throughout the whole image. This signal is attributed to tertiary radiation produced by scattered electrons interacting with the vacuum chamber walls and surrounding material.

This analysis has three advantages:
\begin{enumerate}
 \item The raw measurement of the energy of the signal can help during the alignment procedure.
 \item The identification of the spectrum of background radiation can help to understand its origin and guide experimental shielding improvements. 
 \item Allows an effective background removal.
\end{enumerate}

\begin{figure}[ht]
\includegraphics[width=\columnwidth]{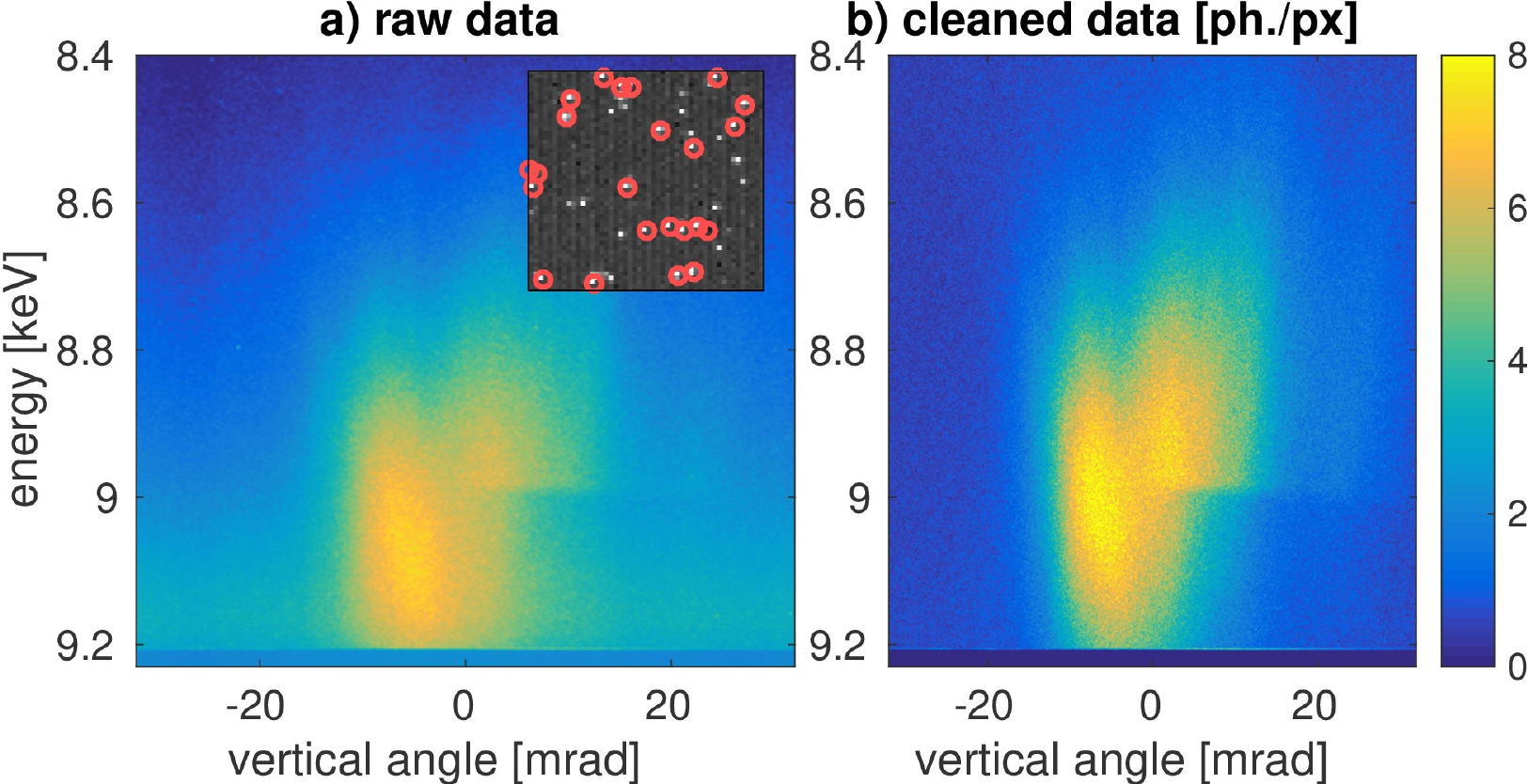}
\caption{\label{rawdata} Raw data from CCD summed over 315 shots (a), the inset shows a single shot image where the 9 keV impacts are highlighted by red circles. (b) image of the same data reconstructed by the single photon counting procedure.}
\end{figure}

Figure \ref{rawdata} demonstrates the effect of this background removal technique. Figure \ref{rawdata} a) shows the raw data digitally accumulated over 315 shots. The inset presents a detail of a single shot image. The algorithm finds those events with energy around 9 keV (shown with red circles) and notes down their positions. The lists of all those positions for all shots in the series are merged and a synthetic data image is constructed from them as a 2D histogram of the impact positions. This is shown in \figref{rawdata} b), where the same data are plotted for comparison in units of photons per pixel. The background which is quite strong especially in the bottom part of the figure has been effectively removed. 

\subsection{Angular resolution}

The angular resolution of the spectrometer is demonstrated in \figref{rawdata}. The energy dispersion direction is vertical on the image and the instrument resolves angularly in the horizontal direction. Half of the beam corresponding to the right part of the figure was propagating through a $3\um$ thick Cu foil while the signal on the left was not obstructed and therefore can be used as a reference beam. The Cu K-edge can be clearly seen as a sharp horizontal line in the right part of the image. 
The uneven shape of the detected signal is caused by the irregularities of the crystal surface. The perfect straightness of the absorption edge however confirms that the effect of mosaicity focusing makes the spectral resolution insensitive to those irregularities.

\begin{figure}[htb]
\includegraphics[width=8cm]{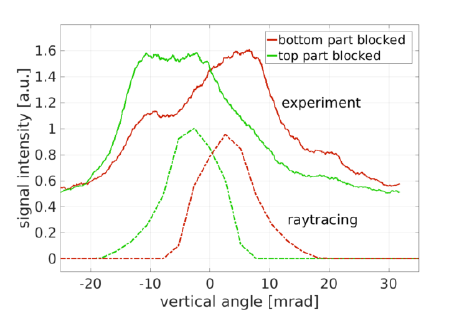}
\caption{\label{horizontal} Horizontal lineouts of the detected signal showing the angular resolution.}
\end{figure}

For a  quantitative analysis of the angular resolution, two series of shots were taken when either one or the other half of the beam was fully blocked.
The horizontal lineouts of the signals integrated over the whole spectral range  are presented in figure \ref{horizontal}. The FWHM of the total signal is 9.2~mm and the difference between the peaks is 3.9 mm. The dashed line presents the output of the ray-tracing algorithm which qualitatively agrees with the experiment. 
The overall width of the signal is however different. The same width of the signal was obtained when the crystal was illuminated by visible light, showing that this effect is caused by the imprecise surface of the crystal.



\subsection{Spectral resolution}
\begin{figure}[htb]
\includegraphics[width=\columnwidth]{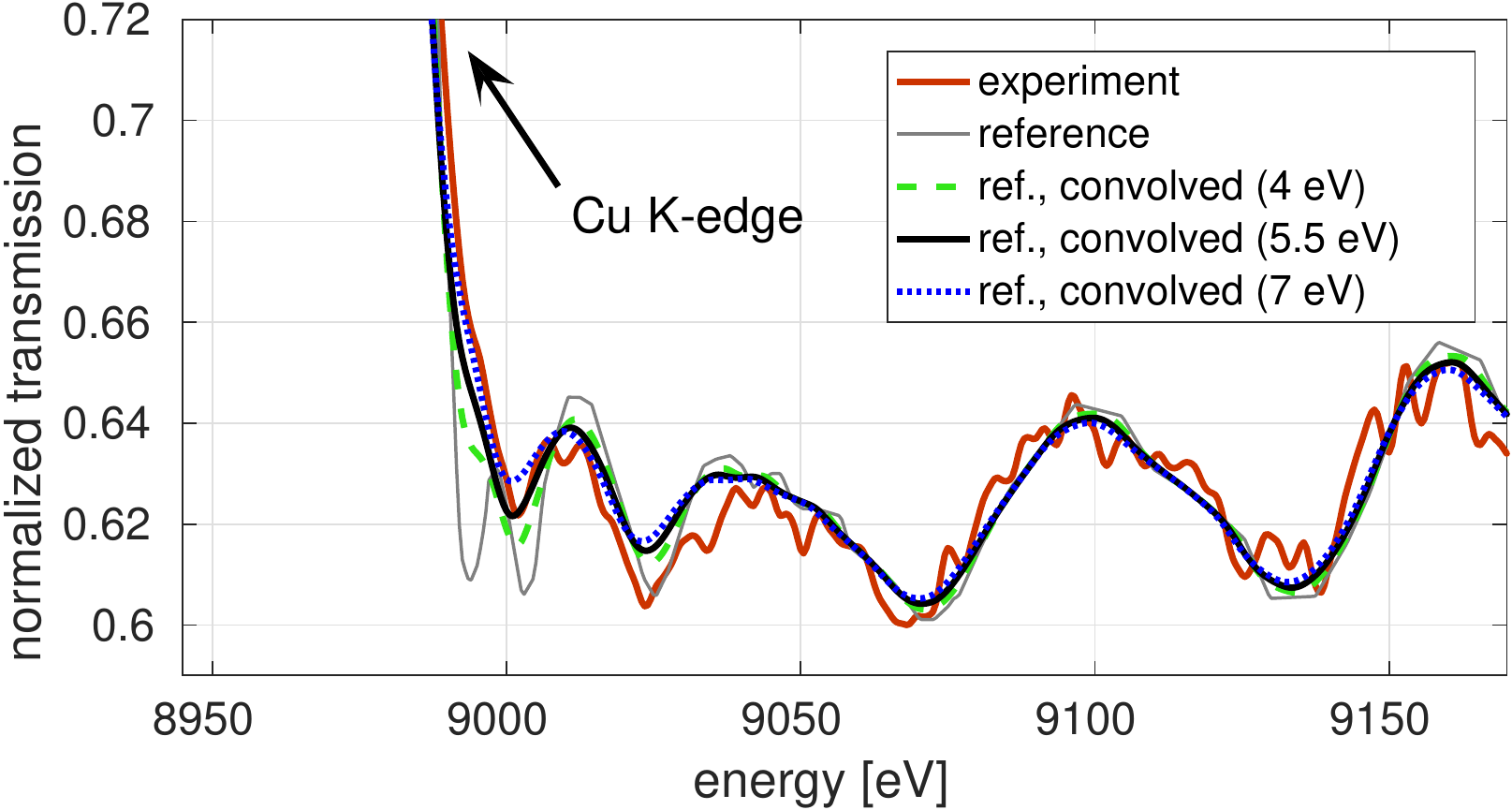}
\caption{\label{final} Measured XANES spectra of $3\um$ thick Cu foil compared to a reference spectrum and to the convolution of this reference with PSF with given FWHM.}
\end{figure}

The experimentally observed absorption spectrum of $3\um$ thick Cu foil backlit with laser-produced betatron radiation can be seen in \figref{final}. 225 laser shots were accumulated, each shot was saved in single CCD datafile and separately processed. The obtained positions of single photon events were gathered and used to plot the profile. The precise energy calibration is done by using the Cu K-edge.

The reference spectrum of polycrystalline Cu \cite{alba,brookhaven} is plotted as a thin gray line.
It has been convolved with the ray-tracing modeled PSF (see \figref{rigaku}) horizontally stretched to various FWHM.
Results for 4, 5.5 and 7 eV FWHM are shown. The spectrum convolved with 5.5 eV wide PSF was selected as the best fit especially due to the agreement of the absorption feature at 9000~eV. Therefore, the FWHM resolution of the spectrometer is estimated to be $\approx$ 5.5 eV.

The PSF has strongly asymmetrical non-Gaussian shape with relatively sharp peak.  The effective resolution can therefore be better for most applications, like observation of narrow features or estimation of spectral line positions. 
Another definition states that resolution is a wavelength of a sinusoidal signal which is transmitted through the system  with 10\% amplitude \cite{dsp}. Numerical analysis of the PSF have shown that the spectral resolution is 4.8 eV when this definition is applied to the 5.5 eV FWHM resolution.


\section{Conclusions}

We have designed and tested a novel HOPG spectrometer in von Hamos geometry optimized for effective use of low flux highly collimated beams. We have measured the XANES spectra of $3\um$ thick Cu foil backlit by a betatron radiation from a LWFA driven by 800 mJ laser and the spectral peaks showing the crystallite structure of the material were resolved. Ray-tracing calculations show that 1.6\% of photons generated by the source in 20~mrad cone are detected. This allowed $33000$ photons on the detector in the range 8.6 -- 9.2 keV in each shot. Accumulation of $\approx 200$ shots was sufficient to obtain enough photons to detect the narrow, low intensity XANES peaks. A separate experiment using a Cu \ka~ source confirmed the strongly asymmetrical PSF of the spectrometer as was predicted by the ray-tracing code. By comparing the measured spectra to reference data \cite{alba,brookhaven} the spectral resolution we are able to infer was 4.8~eV at the wavelength of Cu K-edge (9~keV).
We have demonstrated the angular resolution of the spectrometer which can be used to acquire both reference and data spectra at the same time, allowing accurate measurements of transmission.
The developed ray-tracing code predicts the parameters of the spectrometer in various configurations and explains their behavior. This can help with adjustment of the setup for different experiments.

This spectrometer will be used in future x-ray absorption spectroscopy experiments on WDM targets backlit by LWFA produced betatron radiation. It is expected that a single shot spectrum could be obtained if the acceleration would be driven by $\approx 10$~J laser system.

\section{Acknowledgment}
The authors would like to thank to prof. David Riley from the Queen's University Belfast for lending the crystal and to the staff of Rigaku Innovative Technologies Europe s.r.o. for the calibration x-ray source.
The research leading to these results has received funding from 
LASERLAB-EUROPE (grant agreement no. 654148, European Union's Horizon 2020 research and innovation programme), proposal no. LLC002259 and was supported by the project ELI - Extreme Light Infrastructure – phase 2 (CZ.02.1.01/0.0/0.0/15\_008/0000162 ) from European Regional Development Fund. This project has received funding from the European Research Council (ERC) under the European Union's Horizon 2020 research and innovation programme (grant agreement no. 682399). We acknowledge the support of the Swedish Research Council, the Crafoord Foundation and the Knut and Alice Wallenberg Foundation.

\bibliography{rsi-smid}

\end{document}